\documentclass[10pt,twocolumn,letterpaper]{article}

\usepackage{iccv}
\usepackage{times}
\usepackage{epsfig}
\usepackage{graphicx}
\usepackage{amsmath}
\usepackage{amssymb}
\usepackage{float}
\usepackage[utf8]{inputenc}

\usepackage[breaklinks=true,bookmarks=false]{hyperref}

\iccvfinalcopy % *** Uncomment this line for the final submission

\def\iccvPaperID{****} % *** Enter the ICCV Paper ID here

\ificcvfinal\fi
\begin{document}

%%%%%%%%% TITLE
\title{Deep Perceptual Compression}

\author{Yash Patel$^{2}$ , Srikar Appalaraju$^{1}$, R. Manmatha$^{1}$\\
$^{1}$Amazon\\
$^{2}$Center for Machine Perception, Czech Technical University, Prague, Czech Republic \\
{\tt\small patelyas@cmp.felk.cvut.cz, \{srikara,manmatha\}@amazon.com}
}

\maketitle

\begin{abstract}
Several deep learned lossy compression techniques have been proposed in the recent literature. Most of these are optimized by using either MS-SSIM (multi-scale structural similarity) or MSE (mean squared error) as a loss function. Unfortunately, neither of these correlate well with human perception and this is clearly visible from the resulting compressed images. In several cases, the MS-SSIM for deep learned techniques is higher than say a conventional, non-deep learned codec such as JPEG-2000 or BPG.  However, the images produced by these deep learned techniques are in many cases clearly worse to human eyes than those produced by JPEG-2000 or BPG.

We propose the use of an alternative, deep perceptual metric, which has been shown to align better with human perceptual similarity. We then propose Deep Perceptual Compression (DPC) which makes use of an encoder-decoder based image compression model to jointly optimize on the deep perceptual metric and MS-SSIM. Via extensive human evaluations, we show that the proposed method generates visually better results than previous learning based compression methods and JPEG-2000, and is comparable to BPG. Furthermore, we demonstrate that for tasks like object-detection, images compressed with DPC give better accuracy.

\end{abstract}

%-------------------------------------------------------------------------
\section{Introduction}
\label{sec:introduction}
Image compression takes advantage of the redundancy of information in an image to reduce its size. Both the storage and network bandwidth required for the image are reduced by compression and for large datasets the savings can be large. While there are lossless compression formats such as  PNG \cite{boutell1997png}), the bigger reductions are obtained using lossy compression formats such as
JPEG \cite{wallace1992jpeg}, JPEG2000  \cite{skodras2001jpeg} or BPG \cite{bpg}). These lossy formats are handcrafted (not learned). While learned image compression from data using neural networks is not new \cite{munro1989image,jiang1999image,luttrell1988image}, there has recently been a resurgence of  deep learning based techniques for solving this problem\cite{balle2016end,balle2018variational,mentzer2018conditional1,rippel2017real,lee2018context}. These compression schemes often consist of an encoder-decoder network. The loss function usually trades-off distortion and the bit rate \cite{shannon1948mathematical}. The encoder creates a
latent embedding from the image. With this embedding as input and a combination of a quantizer and an entropy coder generates a compact bit-stream for storage. For decompression, the entropy coding is reversed to produce an embedding which is then fed into a decoder to give a reconstructed approximate image as output.
%Please note, quantizer operation is lossy. 

To evaluate the quality of the reconstructed image with respect to the original, measures such as structural similarity (SSIM) \cite{wang2004image} or PSNR (a function of MSE - mean squared error) have been proposed in the past. In recent work, multi-scale structure similarity (MS-SSIM) \cite{wang2003multiscale} has become more popular. PSNR and MS-SSIM were originally formulated as perceptual metrics but don't seem to completely capture certain type of distortions created by learned compression methods.

\begin{figure*}
	\centering
	\includegraphics[width=\textwidth]{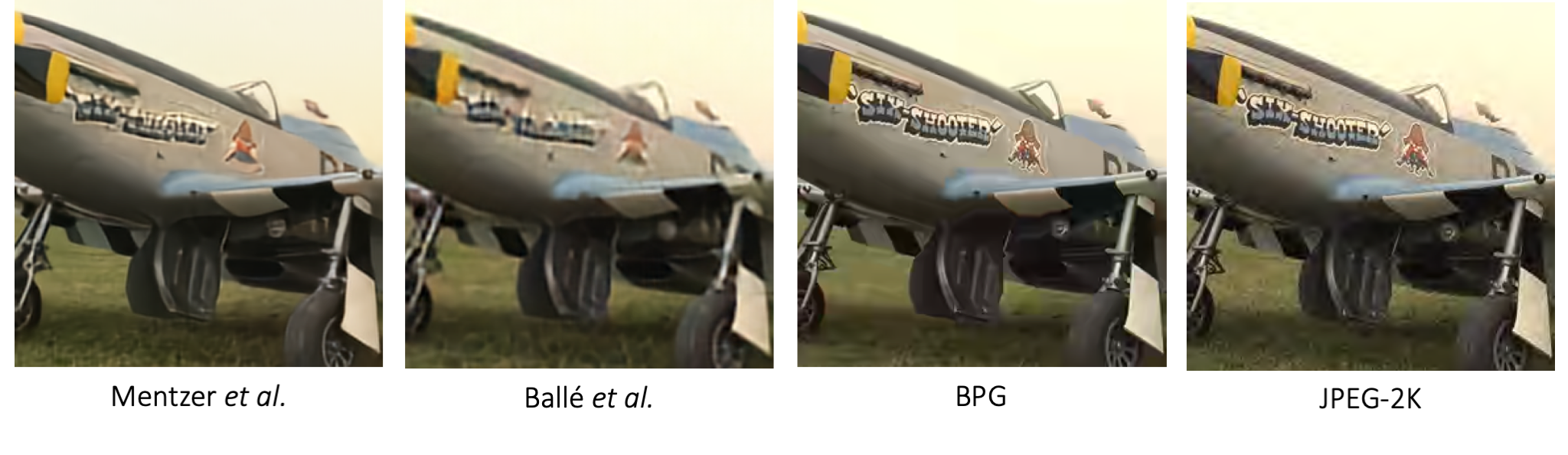}
	\caption{An example from the Kodak dataset. In order of MS-SSIM values Mentzer ~\etal \cite{mentzer2018conditional1} $>$ Ballé  ~\etal \cite{balle2016end} $>$ BPG \cite{bpg} $>$ JPEG-2000 \cite{skodras2001jpeg}. However, visually the foreground and text in BPG and JPEG-2000 are clearly better in quality. Best viewed in color. }
	\label{fig:kodak_samples}
\end{figure*}

Researchers have naturally tried to directly optimize on measures such as MS-SSIM  or MSE (i.e. PSNR) by using it as a loss function \cite{mentzer2018conditional1,rippel2017real,johnston2017improved,balle2018variational,lee2018context}.  The choice of whether PSNR or MS-SSIM is used for evaluation dictates which loss function is used since optimizing for the evaluation metric ensures that the technique achieves a high number on it and a lower number on the other metric. Several claims have been made that such approaches are better compared to engineered compression formats due to their higher MS-SSIM but as Figure \ref{fig:kodak_samples} shows this  misleading. From left to right we have four different techniques ranked in descending order of MS-SSIM values. It is clear that the first two images have many more artifacts than the last two images (the text is not readable in the first two images). Clearly, PSNR and MS-SSIM scores do not reflect image quality or  human perception well. While such scores may be reasonable for measuring engineered codecs which cannot directly optimize these measures, deep learning techniques can directly optimize such metrics leading to this situation. %The metrics themselves leave something to be desired since MS-SSIM cannot differentiate between a (locally) blurry patch and a patch which isn't blurry so this could be one reason for this difference.
The work done by  \cite{anonymous-cite-cvprw-paper} experimentally proves (using human evaluation) that this problem is indeed not confined to one image but occurs across different image compression datasets.

This paper proposes deep perceptual compression (DPC) - a deep learning approach for image compression which uses a Learned Perceptual Image Patch Similarity (LPIPS) metric \cite{zhang18} (deep perceptual metric) as a loss function. Zhang et al \cite{zhang18} use a CNN to compute this metric. Since a CNN in general computes a function, we use their  CNN to compute the deep perceptual loss metric. This perceptual metric was trained by Zhang et al \cite{zhang18} on user judgments on distorted images. To regularize this network we combine the deep perceptual metric with an MS-SSIM loss in a multi-task learning setup (Figure \ref{fig:overall_method}) and train the network end-to-end.

Image generation models with deconvolution based up-sampling are known to generate certain checkerboard patterns with some losses as reported in \cite{sajjadi2017enhancenet,odena2016deconvolution}. In an attempt to minimize checkerboard patters in the reconstructed images, we set the deconvolution up-sampling in a way that kernel sizes are divisible by strides to avoid overlap issue (see \cite{odena2016deconvolution} for a detailed explanation of checkerboard pattern problem).

We show that DPC is better (as judged by humans) than a couple of deep learning techniques \cite{mentzer2018conditional1,balle2016end} as well as JPEG-2000 at a number of bit-rates by doing experiments on several standard compression datasets \footnote{Since we use human judgments, and therefore, require images from each technique for all datasets we were constrained to using deep learning techniques for which the researchers made models available and for this we are thankful}. DPC is better than BPG at some bit-rates while BPG is better at others. Since humans are more sensitive to certain compression artifacts as compared to others, as an alternative to human judgments, we take a pre-trained object detector (ResNet-101) on the COCO-dataset and run it on the images output by each compression algorithm. Absent fine-tuning all algorithms cause some degradation in the object detector performance but DPC suffers the least degradation while at higher bit-rates BPG comes close.
 
%Humans are more sensitive to certain compression artifacts as compared to others. 

The rest of the paper is structured as follows. In Section \ref{sec:related_work}, related work is reviewed. In Section \ref{sec:deep_perceptual_compression}, the proposed method is described and in Section \ref{sec:experiments} experiments and results are discussed. The paper is concluded in Section \ref{sec:conclusions}.

\begin{figure*}[h!]
	\includegraphics[width=\textwidth]{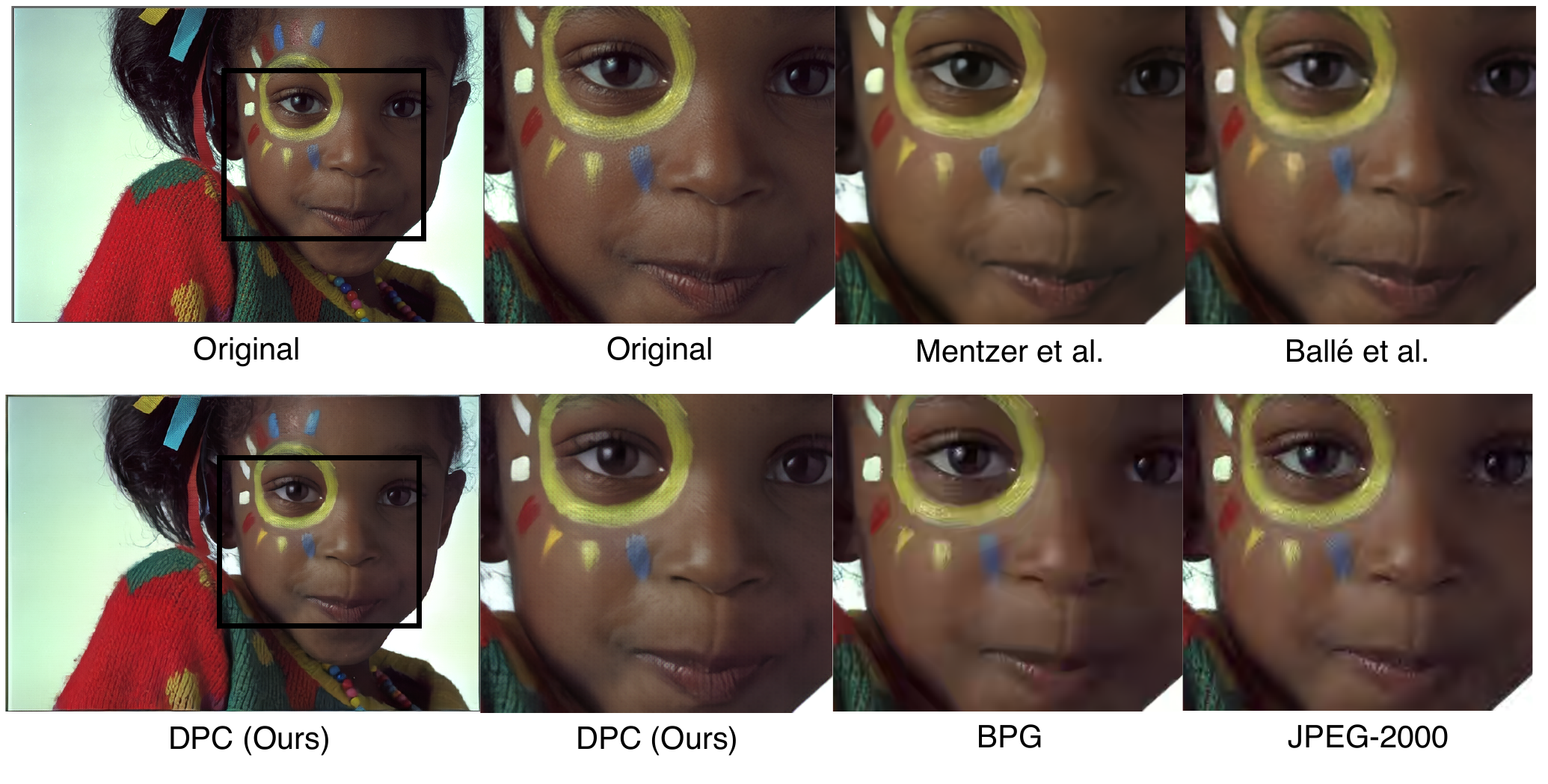}
	\caption{An example from Kodak dataset compressed using different techniquess. Note, DPC (ours) is sharper. Best viewed in color.}
	\label{fig:kodak_samples_2}
\end{figure*}

\section{Related Work}
\label{sec:related_work}

We discuss some related work on learned image compression and perceptual image quality. Many compression models use autoencoders. One difference between models is how the entropy of the data is learned. The entropy model is jointly trained with the encoder and decoder with a rate-distortion trade-off as a loss function \cite{shannon1948mathematical} i.e $L = \beta R + \alpha D$ ($R$ is the rate and $D$ is distortion)). To learn optimal R for a particular D, some have used a fully factorized entropy model \cite{balle2016end,theis2017lossy}, others use context in the quantized space to improve compression using auto-regressive \cite{oord2016pixel} approaches  \cite{mentzer2018conditional1,rippel2017real,li2018learning}. \cite{minnen2018joint} jointly use factorized and auto-regressive approaches to learn entropy.

Apart from the innovation in entropy modeling, some papers have improved on the encoder decoder architectures.  \cite{balle2015density} use GDN activation instead of RELU, \cite{theis2017lossy} adopt ideas from super-resolution work to use pixel-shuffle \cite{shi2016real} in the decoder to better reconstruct the image. \cite{santurkar2018generative,rippel2017real,agustsson2018generative} use adversarial training. On metrics such as MS-SSIM or PSNR they do better than JPEG \cite{wallace1992jpeg}, JPEG2000 \cite{skodras2001jpeg} and some do better than BPG \cite{bpgbellard}. However, as we discussed in the introduction these metrics are misleading.

In the super-resolution literature \cite{ledig2017photo,johnston2017improved}, it has been shown that comparing activations obtained from a  VGG-16 \cite{simonyan2014very} network trained for the classification task on ImageNet \cite{russakovsky2015imagenet} may be used as a perceptual loss function. In other contexts, this approach has been used for neural style transfer \cite{gatys2016image}, for conditional image synthesis \cite{chen2017photographic, dosovitskiy2016generating}. Recently, Zhang et al.\cite{zhang2018unreasonable,chinen2018towards} investigate the effectiveness of these deep CNN's as a perceptual similarity metric. They first show humans a triplet of images which include two distorted versions of an image patch  and the original patch  and ask which  distorted patch is closer to the original. They create a net where the feature responses of standard CNN architectures such as AlexNet \cite{krizhevsky2012imagenet} or VGG-16 \cite{simonyan2014very} (pre-trained on ImageNet) are fed to layers which learn to output distance metrics which reflect the low-level human judgments.

In our work we show that using the deep perceptual metric as a loss function leads to improved image compression results as judged by humans.  We do need to regularize this with an MS-SSIM loss.

%as humans are the best lossy image compressors\cite{bhown2018humans}

\section{Deep Perceptual Compression}
\label{sec:deep_perceptual_compression}

\subsection{Compression Model}
\label{sec:network_architecture}

\begin{figure*}[h!]
	\includegraphics[width=\textwidth]{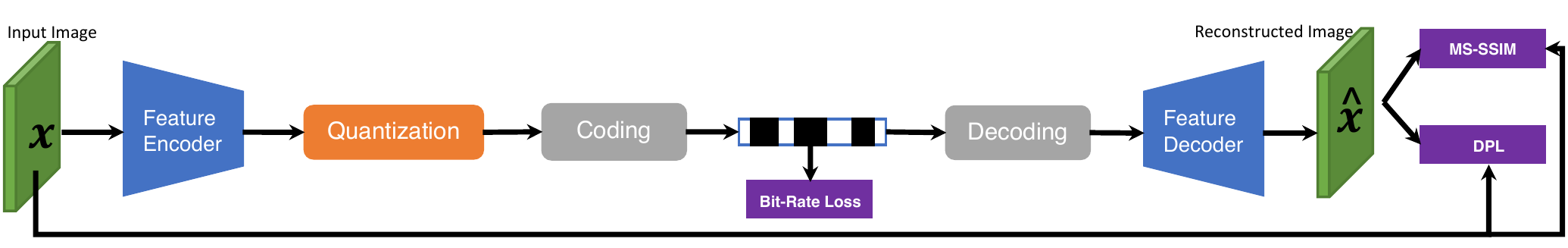}
	\caption{High-level illustration of our image compression model. An input image is fed to the feature encoder, the obtained activations are quantized and entropy-coded. During decompression, these steps are reversed and the decoder outputs the reonstructed image. Purple boxes indicate loss functions, optimization is done  jointly on the bit-rate loss and the two distortion losses (DPL and MS-SSIM).} %In our rate-distortion loss function $L = \beta R + \alpha D$; distortion$D$ is computed by a combination of DPL + MS-SSIM, where as rate$R$ is computed using softmax-cross-entropy. }
	\label{fig:overall_method}
\end{figure*}

We adapt the architecture of Mentzer ~\etal \cite{mentzer2018conditional1} with certain essential modifications to optimize on deep perceptual loss. We explicitly keep certain components (such as quantization and entropy coding) of the original approach \cite{mentzer2018conditional1} to investigate the effect of the proposed deep perceptual loss. An auto-encoder framework is used consisting of stacked residual blocks for the encoder and decoder. In the bottleneck, a quantizer is used for lossy data transformation and an auto-regressive \cite{oord2016pixel} entropy model is used for estimating the probability distribution in the quantized space.
Formally, the compression model consists of an Encoder $E_\theta$, a Decoder $D_\psi$, a Quantizer $Q_c$ and an Entropy coding model $Ent_\gamma$; where $\theta$ and $\phi$ are the learnable parameters represented by a deep residual neural network, $c$ is the number of centers for the lossy quantizer and $\gamma$ is a learnable parameters for the entropy model represented by a 3D pixel-CNN \cite{oord2016pixel}. All these modules are trained and optimized jointly on a Rate-Distortion loss. Please see Fig.\ref{fig:overall_method} for a high-level illustration of the model architecture. 

\begin{figure*}
	\includegraphics[width=\textwidth]{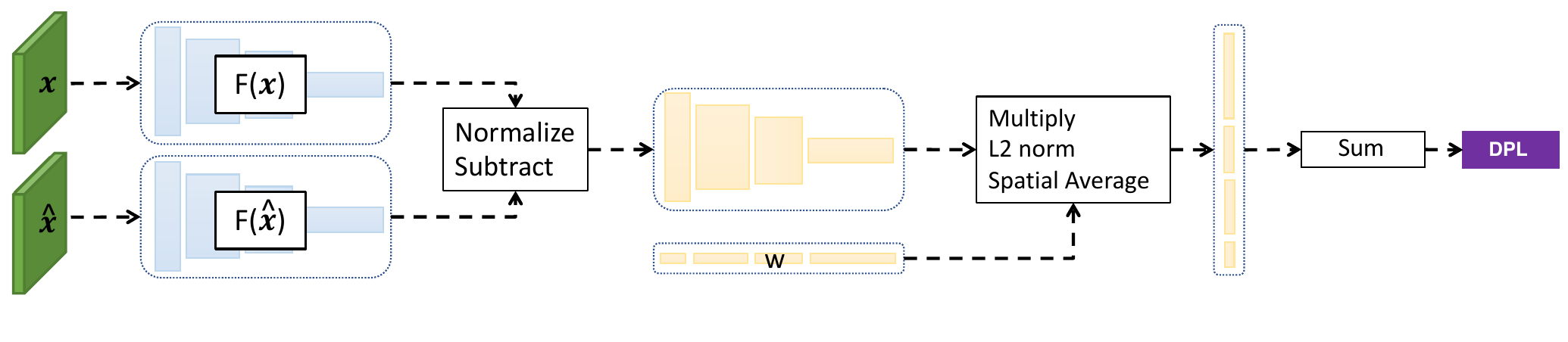}
	\caption{Deep Perceptual Loss: To compute perceptual similarity distance between the original $x$ and recontructed $\hat{x}$ images - first  the deep embeddings $F(x)$ and $F(\hat{x})$ are computed for both, normalized along the channel dimensions, scale each channel by vector $w$ (learned on perceptual similarity dataset), and take the $\ell_{2}$ norm. Finally average across spatial dimensions and sum across channels.}
	\label{fig:dpl}
\end{figure*}

\textbf{Encoder} is a function that takes an image $x$ and computes $E_\theta : \mathbb{R^N} \rightarrow \mathbb{R^M}$  where $N>M$. i.e. for an input image $x$, we get a float-point latent representation $q = E(x)$ where $q$ is a point in a $M$ dimensional space. Next, the \textbf{Quantizer} $Q_c$ discretizes $q$ to $\hat{q}$. It's a bounded discrete space with $c$ centers. Note, this is a lossy transformation. In this work we adopt the differentiable soft-quantization idea from \cite{theis2017lossy,agustsson2017soft}. We use nearest neighbor assignments to compute $ \hat{q}_{i}=Q\left(q_{i}\right) :=\arg \min _{j}\left\|q_{i}-c_{j}\right\| $ where $C = {c_1, c_2, ... c_L}  \subset \mathbb{R}$. In our model we use $c=6$ centers.

\textbf{Entropy model:} Furthermore, $Ent_\gamma$ learns the probability distribution of the quantized co-efficients $\hat{y} = Ent_\gamma(\hat{q})$. $\hat{y}$ is then losslessly encoded into a binary bit-stream using arithmetic coding. In this work (and in \cite{mentzer2018conditional1}), a variant of an auto-regressive model called PixelCNN\cite{oord2016pixel} is used which requires that the next quantized value is conditional on previously seen quantized values  $p(\hat{q})=\prod_{i=1}^{m} p\left(\hat{q}_{i} | \hat{q}_{i-1}, \ldots, \hat{q}_{1}\right)$. Softmax-cross-entropy loss is used as the Rate R coding cost. 

%TODO: can we represent inquealities in latex short-form.
\textbf{Importance map:} We also used variable bit-allocation, since in images there is great variability in information content across spatial locations. Specifically, we take the last layer of Encoder $E_\theta$ and add a single-channel 2D output of the form: $y \in \mathbb{R}^{\frac{W}{8} \times \frac{H}{8} \times 1}$. This $y$ is further expanded into a mask $m \in \mathbb{R}^{\frac{W}{8} \times \frac{H}{8} \times K}$ with the same dimensionality as $q$. The following rules determine the values of the map $m$: $\scriptstyle m_{i, j, k}=\scriptstyle \left\{\begin{array}{ll}{1} & {\text { if } k<y_{i, j}} \\ {\left(y_{i, j}-k\right)} & {\text { if } k \leq y_{i, j} \leq k+1} \\ {0} & {\text { if } k+1>y_{i, j}}\end{array}\right.
$ where $y_{i,j}$ denotes the value at $(i,j)$. This mask is then point-wise multiplied with $q$ i.e.  ${q} \leftarrow {q} \odot\lceil{m}\rceil$ to give a spatially adaptive quantized feature map $q$. Please refer to the original work \cite{mentzer2018conditional1} for more specifics.

Finally, the \textbf{Decoder} reconstructs the quantized latent vector $\hat{q}$ back to an image, $D_\psi: \mathbb{R^M} \rightarrow \mathbb{R^N}$ i.e. $\hat{x} = D_\psi(\hat{q})$. The goal is to learn a compact quantized latent representation $\hat{q}$ such that the distortion between the original image $x$ and the reconstructed image $\hat{x}$ is minimum. This is achieved by using a rate-distortion based loss function i.e.  $L = \beta rate R + \alpha distortion D$. In Mentzer et al \cite{mentzer2018conditional1}, MS-SSIM\cite{wang2003multiscale}  is used for measuring distortion between images $D$. 

\textbf{Checkerboard patterns}: Image generation models with deconvolution based up-sampling are known to generate certain checkerboard patterns depending on the loss function. A number of proposals have been made
to solve them\cite{sajjadi2017enhancenet,odena2016deconvolution}. To minimize checkerboard patters in the reconstructed images, we set deconvolution up-sampling in a way that kernel sizes are divisible by strides to avoid overlap issue. We refer the readers to \cite{odena2016deconvolution} for more on the checkerboard pattern problem. Specifically, we use we kernels of size 2 and stride 2 in the Decoder. 

%Specifically, input images $x$ are random crops of size $(batchsize, 160,160,3)$. The $E_\theta$ and Quantizer $Q_c$ transformation gives us a quantized latent vector of dimension $(32, 20, 20, 6)$.

The combination of the above mentioned modules match the state-of-the-art compression performance on Kodak dataset on MS-SSIM metric (circa CVPR 2018). Newer work since then has shown some improvements on MS-SSIM or PSRN (\cite{balle2018variational,minnen2018joint,lee2018context}). However, as we pointed out in the introduction MS-SSIM is not a good evaluation metric. We will instead use human judgments and these models are not available for generating images across several datasets for human judgments.  We now discuss the internals of Perceptual loss and its effects on lossy image compression.

\subsection{Deep Perceptual Loss}
\label{sec:deep_perceptual_loss}

Zhang et al. \cite{zhang18} show the utility of deep CNNs to measure perceptual similarity. It has been observed that comparing internal activations from deep CNNs such as VGG-16 \cite{simonyan2014very} or AlexNet \cite{krizhevsky2012imagenet} acts as a better perceptual similarity metric than MS-SSIM or PSNR. We use the deep perceptual metric for both training and as one of the evaluation metric on test data. We make use of activations from five $ReLU$  layers after each $conv$ block in the VGG-16 \cite{simonyan2014very} architecture with batch normalizations.

Feed-forward is performed on the VGG-16 for both the original ($x$) and reconstructed images ($\hat{x}$). Let $L$ be the set of layers used for loss calculation (five for our setup) and, a function $F(x)$ denotes feed-forward on an input image $x$. $F(x)$ and $F(\hat{x})$ return two stacks of feature activations for all $L$ layers. The deep perceptual loss is then computed as follows:
\begin{itemize}

\item $F(x)$ and $F(\hat{x})$ are unit-normalized in the channel dimension. Let us call these, $z^{l}_{x}, z^{l}_{\hat{x}} \in \mathbb{R}^{H_{l} \times W_{l} \times C_{l}}$ where $l \in L$. ($H_{l}, W_{l}$ are the spatial dimensions of the given activation map and $C_{l}$ is the number of channels).
\item $z^{l}_{x}, z^{l}_{\hat{x}}$ are scaled channel wise by multiplying with the vector $w^{l} \in \mathbb{R}^{C_{l}}$ 
\item The $\ell_{2}$ distance is then computed and an average over spatial dimensions are taken.
\item Finally, a channel-wise sum is taken, outputing the deep perceptual loss.
\end{itemize}

Equation. \ref{eqn:dpl}  and Figure. \ref{fig:dpl} summarize the  deep perceptual loss computation. Note that the weights in $F$ are learned for image classification on the ImageNet dataset \cite{russakovsky2015imagenet} and are kept fixed. $w$ are the linear weights learned on top of $F$ on the Berkeley-Adobe Perceptual Patch Similarity Dataset \cite{zhang2018unreasonable}. Note we use the trained model provided by Zhang et al. \cite{zhang18} to compute the loss.

\begin{equation}
DPL(x,\hat{x}) = \sum_{l} \frac{1}{H_{l}W_{l}} \sum_{h,w} || w_{l} \odot  (z^{l}_{\hat{x},h,w} - z^{l}_{x,h,w}) ||_{2}^{2}
\label{eqn:dpl}
\end{equation}

For training we make use of MS-SSIM for regularization, resulting in the final distortation loss to be: $D(x, \hat{x}) = DPL(x,\hat{x}) + \lambda MSSSIM(x,\hat{x})$. In practice, we use $\lambda=1$ in our training setup.

\subsection{Training Details}
\label{sec:training_details}

We make use of the Adam optimizer \cite{kingma2014adam} with an initial learning rate of $4 \times 10^{-3}$ and a batch-size of $30$. The learning rate is decayed by a factor of $10$ in every two epochs (step-decay). The overall loss function is: $L(x,\hat{x}) = \alpha D + \beta R$, where $R$ is the rate loss and $D$ is a distortion loss, which for DPC is a linear combination of the deep perceptual loss (see \ref{sec:deep_perceptual_loss}) and MS-SSIM (weighted equally). Further, similar to \cite{mentzer2018conditional1}, we clip the rate term to $max(t, \beta R)$ to make the model converge to a certain bit-rate, $t$. The training is done on the training set of ImageNet dataset the from Large Scale Visual Recognition Challenge 2012 (ILSVRC2012) \cite{russakovsky2015imagenet}, with the mentioned setup, we observe convergence in six epochs.

By varying the model hyper-parameters such as the number of channels in the  bottleneck, weight for distortion loss ($\alpha$), target bit-rate ($t$), we obtain multiple models in the bit-per-pixel range of $0.15$ to $1.0$. Similarly we reproduce the models for \cite{mentzer2018conditional1,balle2016end} at different bpp values. \footnote{Note that in the case of \cite{balle2016end} we used an MS-SSIM loss instead of MSE loss as was done in the original paper but this does not change the general conclusions of the paper.}

\section{Experiments}
\label{sec:experiments}

%\subsection{Setup for Human Evaluations}
%\label{sec:human_eval}

We extensively evaluate image compression techniques for human perceptual similarity using a  two alternative forced choice (2AFC) approach. 2AFC is a known way of performing perceptual similarity evaluation and has been used by \cite{sajjadi2017enhancenet} for evaluating super-resolution techniques. The study is conducted on the Amazon MTurk platform where an evaluator is show the original image along with the compressed images from two techniques on each side. They are asked to choose the image which is more similar to the original. We show the entire image along with a synchronized (on all three) magnifying glass to observe finer details. This gives them a global context of the whole image and at the same time provides a quick way to access local regions. No time limit was placed for this human experiment.

In this setup, we compare the proposed DPC, two engineered (JPEG-2000 \cite{skodras2001jpeg}, BPG \cite{bpg}) and two learning based (Mentzer ~\etal \cite{mentzer2018conditional1}, Ballé ~\etal \cite{balle2016end}) compression techniques by choosing all possible combinations (ten pairs in total). Further, we do this at four different compression levels - i.e. bits-per-pixel (bpp) values: $0.23, 0.37, 0.67, 1.0$. The study is conducted on four standard datasets: Kodak \cite{kodak}, Urban100 \cite{huang2015single}, Set14 \cite{zeyde2010single} and Set5 \cite{bevilacqua2012low}.

We have a total of $5720$ pairs ($10$ pairs for five methods, $4$ bpp values and $143$ images in total). For each such pair, we obtain $5$ evaluations resulting in a total of $28600$ HITs.

\subsection{Image Compression Results}
\label{sec:results}

\begin{figure*}[h!]
	\centering
	\includegraphics[width=\textwidth]{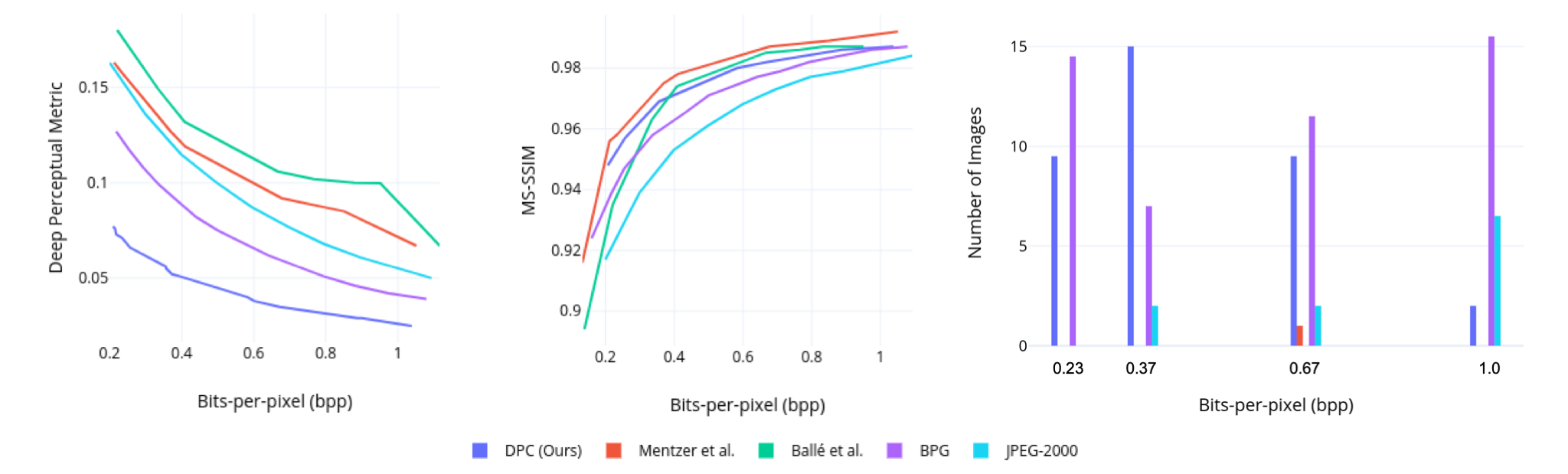}
	\caption{Evaluation on the Kodak dataset \cite{kodak} using the deep perceptual metric (left), MS-SSIM (middle) and human study (right). In this case, Ballé ~\etal is never the best method for any image. Best viewed in color.}
	\label{fig:kodak_plots}
\end{figure*}

\begin{figure*}[h!]
	\centering
	\includegraphics[width=\textwidth]{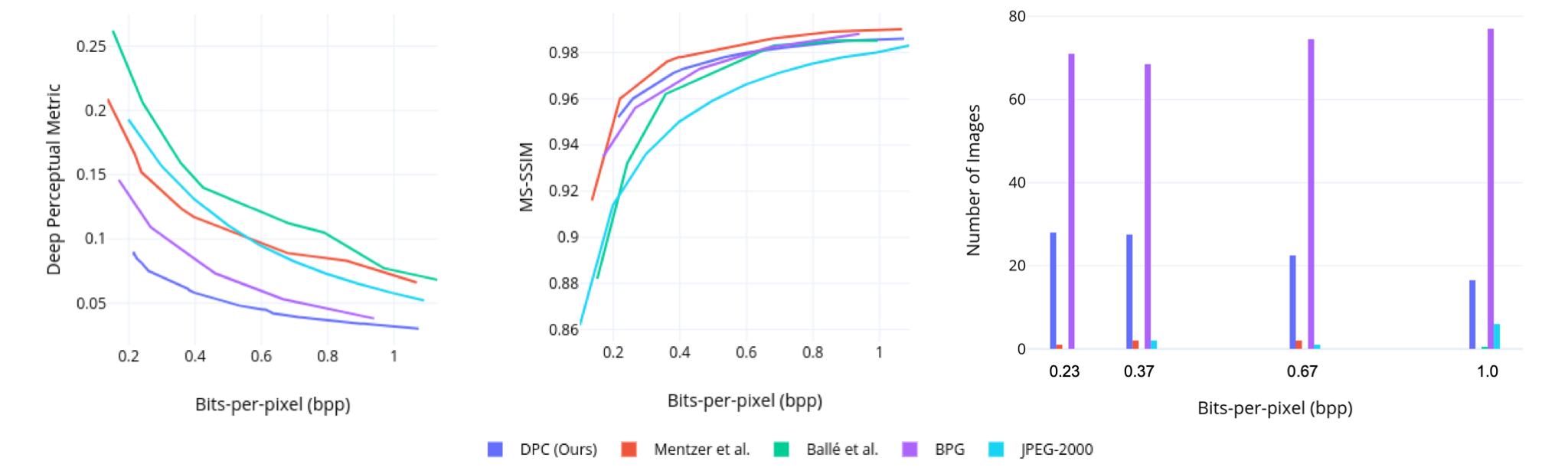}
	\caption{Evaluation on the Urban100 dataset \cite{huang2015single} using the deep perceptual metric (left), MS-SSIM (middle) and human study (right). Best viewed in color.}
	\label{fig:urban100_plots}
\end{figure*}

\begin{figure*}[h!]
	\centering
	\includegraphics[width=\textwidth]{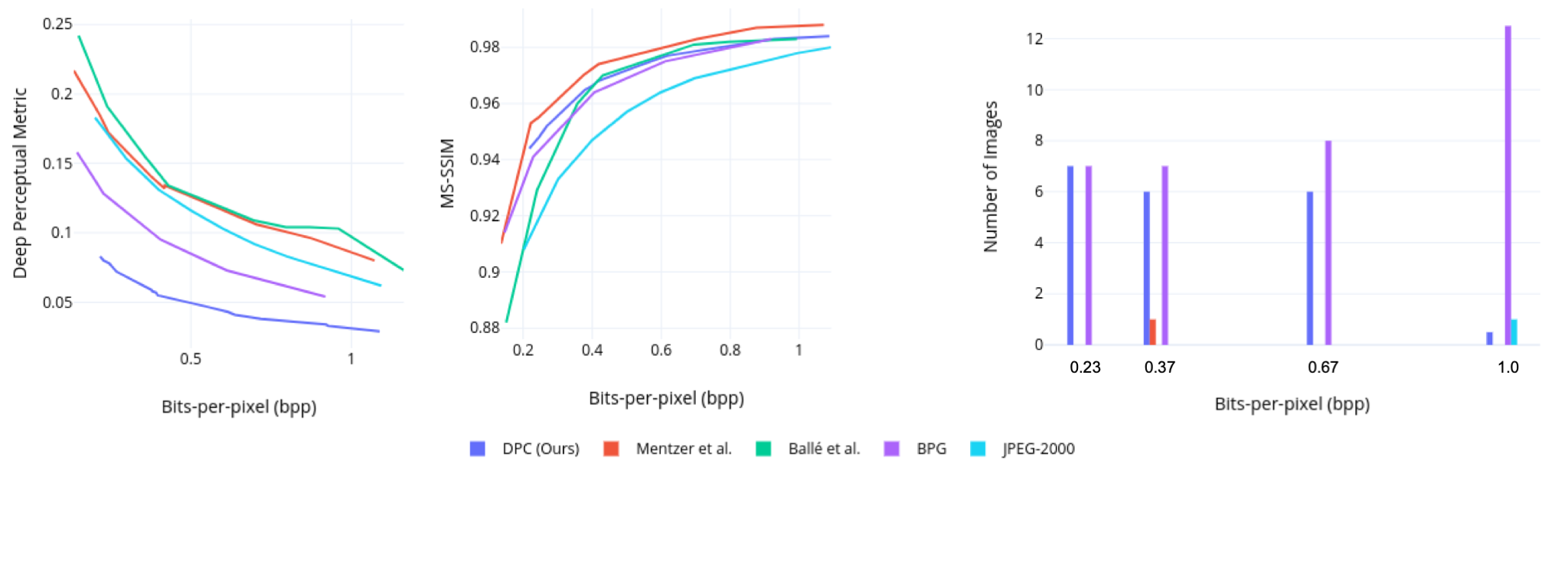}
	\caption{Evaluation on Set14 dataset \cite{zeyde2010single} using the deep perceptual metric (left), MS-SSIM (middle) and human study (right). In this case, Ballé ~\etal is never the best method for any image. Best viewed in color.}
	\label{fig:set14_plots}
\end{figure*}

\begin{figure*}[h!]
	\centering
	\includegraphics[width=\textwidth]{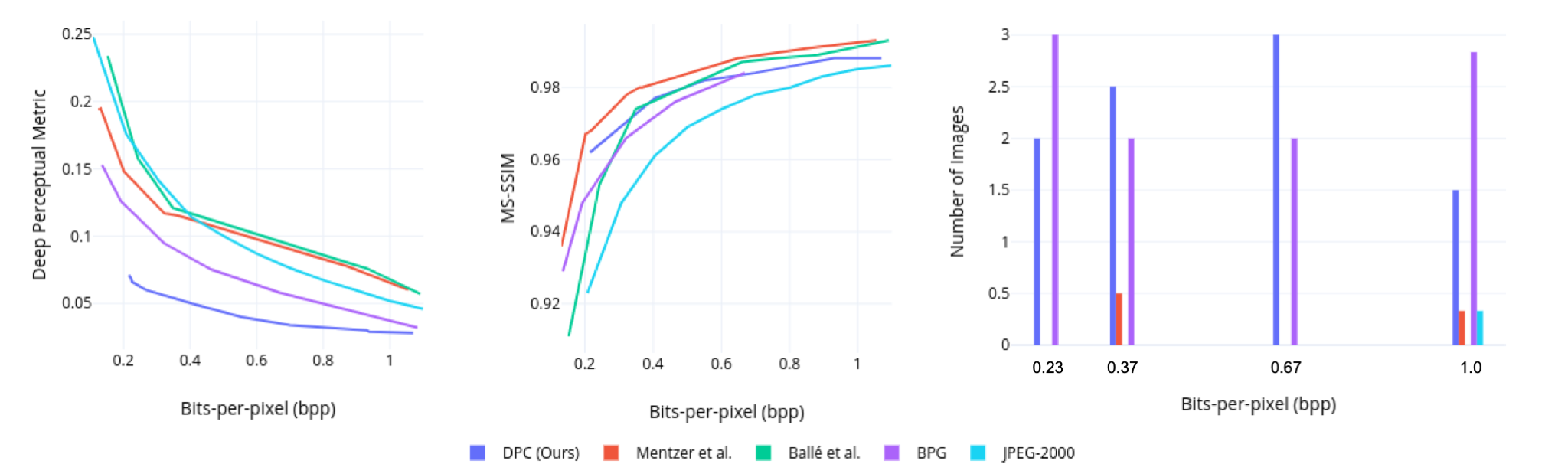}
	\caption{Evaluation on Set5 dataset \cite{bevilacqua2012low} using the deep perceptual metric (left), MS-SSIM (middle) and human study (right). In this case, Ballé ~\etal is never the best method for any image. Best viewed in color.}
	\label{fig:set5_plots}
\end{figure*}

\begin{figure*}[h!]
	\centering
	\includegraphics[width=\textwidth]{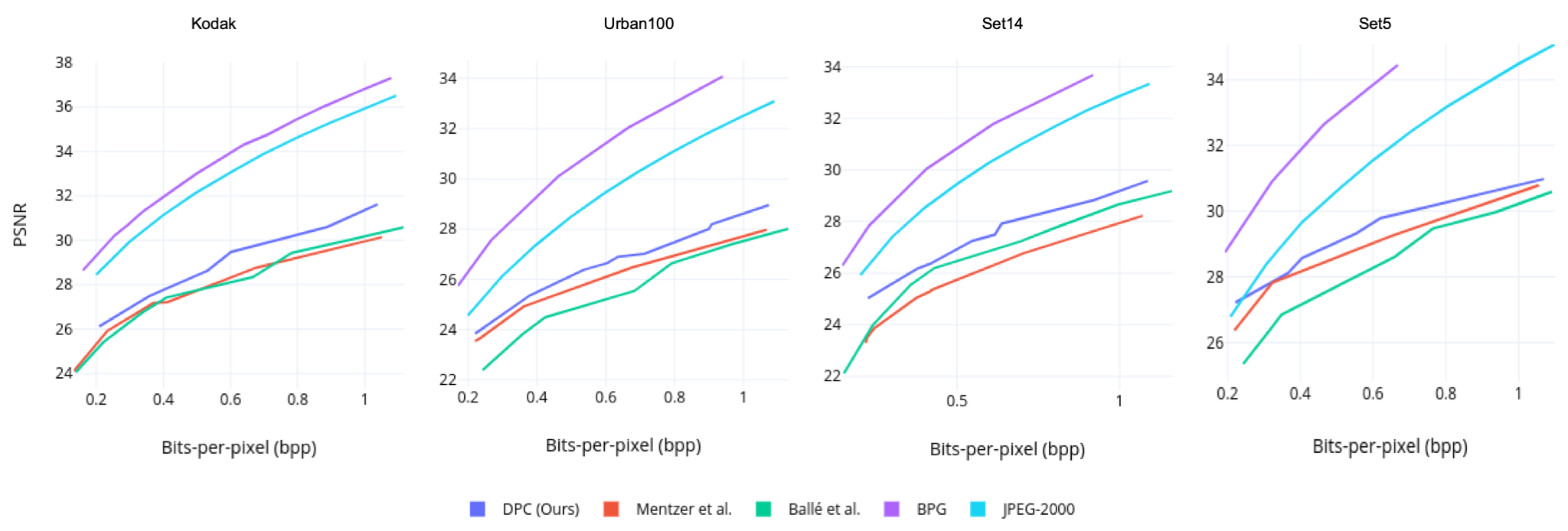}
	\caption{PSNR plots for all four datasets (from left to right): Kodak \cite{kodak}, Urban100 \cite{huang2015single}, Set14 \cite{zeyde2010single} and Set5 \cite{bevilacqua2012low}. Best viewed in color.}
	\label{fig:psnr_plots}
\end{figure*}

\begin{figure*}[h!]
	\centering
	\includegraphics[width=\textwidth]{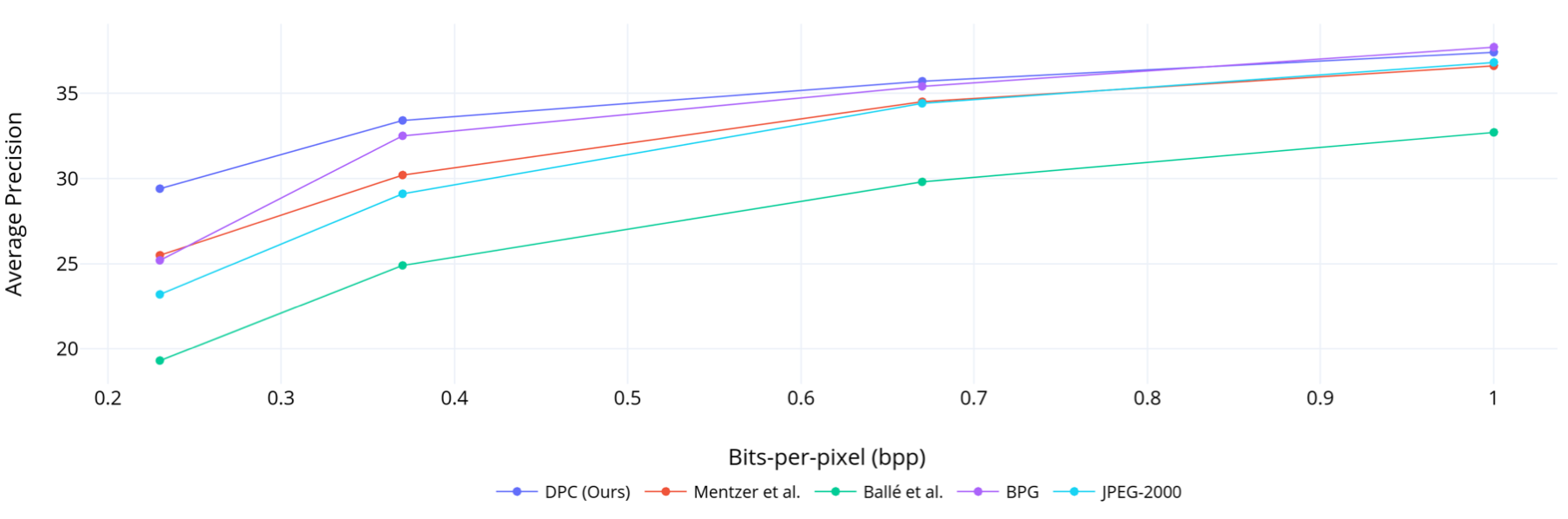}
	\caption{Object detection on the validation set of MS-COCO dataset \cite{mscoco}. The plot shows Average Precision (AP) against varying bits-per-pixel for various compression methods. Best viewed in color.}
	\label{fig:object_detection_mscoco}
\end{figure*}

For each test dataset, we compress all the images using each model. For each model and each bpp we compute an average for all images for each metric. We do the same for different bpp's so that we get multiple points on the deep perceptual metric vs bpp, MS-SSIM vs bpp and PSNR vs bpp curves. We interpolate the values between two such points and we do not extrapolate the values outside the bpp range. Note that for the deep perceptual metric, a lower value is better and for MS-SSIM and PSNR higher is better.

For human evaluation, for each image and a given bpp value we have $5$ pair-wise votes in the form of method-A vs method-B. Since we have all possible pairs for the five methods under consideration, we aggregate these votes and obtain the method which does best for the given image (at a particular bpp value) based on maximum votes. In the Figures \ref{fig:kodak_plots}, \ref{fig:urban100_plots}, \ref{fig:set14_plots} and \ref{fig:set5_plots}, we show the number of images (y-axis) for which a method performs best.

The comparisons for Kodak \cite{kodak} are shown in Figure \ref{fig:kodak_plots}. We observe that Mentzer ~\etal \cite{mentzer2018conditional1} despite the highest MS-SSIM score for all bpp values ranks $4th$ in the human study. Ballé ~\etal \cite{balle2016end} obtains a higher MS-SSIM score compared to DPC, BPG and JPEG-2000 after $0.3$ bpp, despite which it ranks worst among the five. DPC with the lowest deep perceptual metric scores performs better than Mentzer ~\etal \cite{mentzer2018conditional1},  Ballé ~\etal \cite{balle2016end} and JPEG-2000 at all bit-rates. The performance is comparable to BPG and better at $0.37$ bpp value. PSNR plots on all four datasets are shown in Figure. \ref{fig:psnr_plots}, it can be observed that the conventional methods (JPEG-2000, BPG) have significantly higher PSNR scores, although DPC outperforms JPEG-2000 and is comparable to BPG in human study. These observations show that both PSNR and MS-SSIM are inadequate metrices to judge perceptual similarity for learned compression techniques.

Similarly comparisons for other datasets are made in: Figure \ref{fig:urban100_plots} for Urban100 \cite{huang2015single}, Figure \ref{fig:set14_plots} for Set14 \cite{zeyde2010single} and Figure \ref{fig:set5_plots} for Set5 \cite{bevilacqua2012low}.

\subsection{Object Detection Results}
\label{sec:object_detection}

While the compressed images need to be perceptually good, they should also be useful for subsequent computer vision tasks. It was observed by Dwibedi ~\etal \cite{dwibedi2017cut} that for object detectors such as Faster-RCNN \cite{ren2015faster}  region-based consistency is important and pixel level artifacts can significantly affect the performace. In this section, we evaluate  different compression techniques for a subsequent task of object detection on  MS-COCO validation dataset \cite{mscoco}.

We use a pre-trained Faster-RCNN \cite{ren2015faster} model with a $ResNet-101$\cite{he2016deep} based backbone for its relatively high average precision and capability to detect smaller objects. The performance is measured using, average precision (AP), AP is the average over multiple IoU (the minimum IoU to consider a positive match). We use AP@[.5:.95] which corresponds to the average AP for IoU from 0.5 to 0.95 with a step size of 0.05. With the original MS-COCO images, this model attains a performance of $40.1\%$ AP . For each compression method, we compress and reconstruct the image at four different bit-rate values: $0.23, 0.37, 0.67, 1.0$ (same values as used for human evaluation) and then we evaluate them for object detection. The performance of competing compression methods are reported in Figure. \ref{fig:object_detection_mscoco}. It can be clearly seen that at low bit-rates the proposed DPC significantly outperforms the competing methods. At $1.0$ bit-rate, the performance is very close to that of BPG. Please note, as we did not fine-tune networks with the compressed images, there is degradation in performance from the current state-of-the-art.

\section{Conclusions}
\label{sec:conclusions}
We have demonstrated that using a deep perceptual metric as a loss with MS-SSIM as a regularizer one can obtain good image compression as judged by humans on several standard compression datasets. MS-SSIM and PSNR are not good metrics for evaluating image compression and human judgments are more reliable. We also show that DPC compression causes less degradation in a pre-trained object detector than a number of other approaches.

\section*{Acknowledgment}
We would like to thank Joel Chan and Peter Hallinan for helping us in setting up the human evaluations.

{\small
\bibliographystyle{ieee}
\bibliography{example_paper}
}
\end{document}